%% file: main.tex
\documentclass[journal]{IEEEtran}
\usepackage[nolist]{acronym}
\usepackage{amsmath,amsfonts}
\usepackage{algorithmic}
\usepackage{algorithm}
\usepackage{array}
\usepackage[caption=false,font=normalsize,labelfont=sf,textfont=sf]{subfig}
\usepackage{textcomp}
\usepackage{stfloats}
\usepackage{url}
\usepackage{verbatim}
\usepackage{graphicx}
\usepackage{subcaption} 
\usepackage[export]{adjustbox}
\usepackage{wrapfig}
\usepackage{cite}
\usepackage{tabularx} 
\usepackage{array} 
\newcolumntype{Y}{>{\centering\arraybackslash}X}
\usepackage{float}
\usepackage{xcolor}
\usepackage{tabularx}
\usepackage[table]{xcolor} 
\usepackage{pdflscape}


\begin{document}

%

\title{Sovereign AI for 6G: Towards the Future of AI-Native Networks}

\author{\IEEEauthorblockN{Swarna Bindu Chetty, David Grace, Simon Saunders, Paul Harris,  
Eirini Eleni Tsiropoulou, Tony Quek, Hamed Ahmadi}

\thanks{S.B. Chetty, D Grace and H Ahmadi are with the University of York, UK. E. E. Tsiropoulou is with Arizona State University, USA. Simon Saunders is with the University of Bristol UK, P Harris is with VIAVI Solutions. T. Quek is with Singapore University of Technology and Design, Singapore.}
}
\maketitle

\begin{abstract}
 The advent of Generative Artificial Intelligence (GenAI), Large Language Models (LLMs), and Large Telecom Models (LTM) significantly reshapes mobile networks, especially as the telecom industry transitions from 5G's cloud-centric to AI-native 6G architectures. This transition unlocks unprecedented capabilities in real-time automation, semantic networking, and autonomous service orchestration. However, it introduces critical risks related to data sovereignty, security, explainability, and regulatory compliance especially when AI models are trained, deployed, or governed externally. 
This paper introduces the concept of `Sovereign AI' as a strategic imperative for 6G, proposing architectural, operational, and governance frameworks that enable national or operator-level control over AI development, deployment, and life-cycle management. Focusing on O-RAN architecture, we explore how sovereign AI-based xApps and rApps can be deployed Near-RT and Non-RT RICs to ensure policy-aligned control, secure model updates, and federated learning across trusted infrastructure. We analyse global strategies, technical enablers, and challenges across safety, talent, and model governance. Our findings underscore that Sovereign AI is not just a regulatory necessity but a foundational pillar for secure, resilient, and ethically-aligned 6G networks.


\end{abstract}

\section{Introduction}

Early generations of mobile networks focused on providing connectivity between human users and did not consider efficiency of using network resources. In later generations like \ac{LTE}, connecting machines and machine-to-machine communication started to become an important issue. With the emergence of this massive number of users efficient use of network resources became a key factor in mobile networks paving the way to use \ac{AI} in networks. \ac{5G} is considered to be the use case-oriented generation in which the applications like mobile streaming/gaming, autonomous systems and \ac{VR}/\ac{AR} require services that are high data rate, low latency, and/or highly reliable. \ac{AI} and \ac{ML} have been used in \ac{5G} for a range of applications including and not limited to channel prediction, user behaviour monitoring and prediction, network deployment and energy optimisation \cite{andrews_5G_2014}.

The story, however, is going to be different in \ac{6G}. 
It needs to support use cases with more stringent requirements. \ac{6G} will support applications like \ac{DT}, metaverse, and \ac{XR} that demand data rate, reliability and latency beyond what could be provided by \ac{5G} \cite{Ahmadi_DT_21}. 
Also \ac{AI} and \ac{ML} have progressed significantly and now are embedded in all aspects of the modern life. \acp{LLM} and \ac{GenAI} have changed the way almost all systems used to operate and telecommunication networks are not an exception. \ac{6G} and beyond are correctly expected to be the \ac{AI} native generations meaning that \ac{AI} will no longer be an auxiliary tool but will become the backbone of next-generation telecommunication networks, enabling real-time automation, dynamic resource allocation, and cognitive decision-making. It will be present in all stages of design, development, operation and troubleshooting of the network and its components.

Since the early days of using \ac{AI}/\ac{ML} in mobile networks, i.e. Cognitive Radio hype, access to network's real data was a massive challenge for \ac{ML} researchers and start ups. Operators had full control on network's data and correctly have been very protective of that as it involved their network settings and the privacy of their users. However, \ac{AI}/\ac{ML} is changing this trend. Although operators can still have full control of their network's data, they need to train \ac{AI}/\ac{ML} models over their network data to optimise their resource allocation and network performance. Since training and maintaining modern and complicated \ac{ML} models like \acp{LLM} and \acp{LTM} are extremely challenging and costly tasks, for these the operators may relay on \ac{AI}/\ac{ML} service providers \cite{WhitePaper_LTM}. Control over trained models and inferred data will become even more complicated in scenarios where the infrastructure is shared like \acp{NHN} and more sophisticated \ac{ML} technologies like \ac{LLM}, \ac{GenAI}, and Agentic AI are used. 
Here the operators face the dilemma of accepting the costs and challenges of maintaining control over the \ac{AI} versus unknown challenges of embracing unleashed \ac{AI}. 

\begin{figure}
    \centering
    \includegraphics[width=0.5\textwidth]{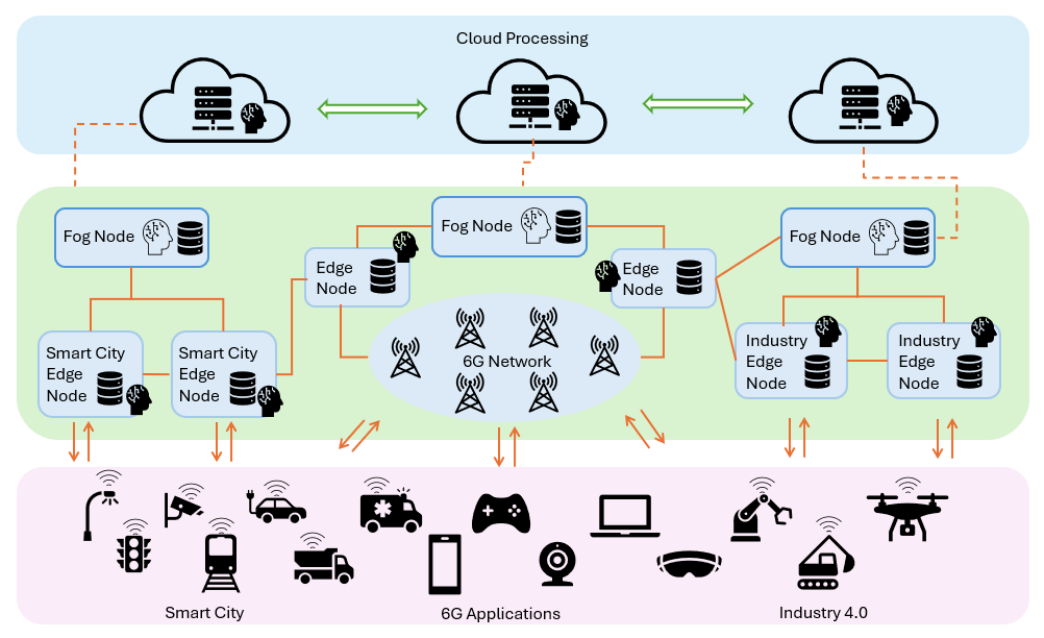}
    \caption{6G AI-Native Architecture: An Intelligent Multi-Tier Ecosystem Enabling Real-Time Applications across Smart Cities and Industry 4.0 via Edge, Fog, and Cloud AI Integration.}
    \label{fig:6g-AI-native-archi}
\end{figure}

The aforementioned challenges highlight the importance and necessity of sovereign \ac{AI} as a strategic response to the challenges introduced by advanced AI technologies in 6G networks. Figure \ref{fig:6g-AI-native-archi} shows the AI-native 6G architecture with distributed intelligence, emphasising the role of Sovereign AI in ensuring secure and autonomous network operations.
In this paper, we explore how modern \ac{AI} techniques, including \ac{GenAI} and \ac{LTM} can be integrated into telecom systems, and examine the implications this holds for trust, governance, and operational autonomy. We introduce the concept of network Sovereign \ac{AI}, detailing its key features and capabilities, and propose practical mechanisms, such as \ac{FL}, synthetic data, and explainable \ac{AI}, to support its implementation within \ac{O-RAN} environments. Finally, we review global strategies and real-world case studies, and discuss the essential role of domestic research, infrastructure, and skilled talent in enabling long-term Sovereign \ac{AI} deployment.


\section{Modern AI}
In the recent years, \ac{AI} has evolved from a futuristic concept into a driving force revolutionising how industries operate, including telecommunications. By enabling machines to reason, learn, and act autonomously, \ac{AI} now underpins the cognitive infrastructure required for next-generation networks. 

Conventionally \ac{ML} is categorised as Supervised Learning,  Unsupervised Learning and \ac{RL}.
To address the challenges of learning from large sets of data with unknown set of features \ac{DL} has been proposed. \ac{DL} utilises multi-layered \acp{NN}, inspired by the architecture of the human brain, to model highly complex and non-linear relationships in data. While \ac{DL} has been mainly developed for supervised and unsupervised learning, it has also been combined with \ac{RL} leading to extending capabilities of \ac{RL}. Advances in \ac{DRL} and introduction of techniques like actor-critic and on-policy, off policy methods paved the way to further advancements in \ac{ML} and dawn of \ac{GenAI}.

\subsection{Generative AI and Large Language Models}
Ushering in a new wave of \ac{AI} innovation, \ac{GenAI} represents a fundamental shift from traditional analytical and predictive paradigms towards autonomous content creation. Rather than passively interpreting data, \ac{GenAI} systems actively generate text, images, audio, code and other modalities, often indistinguishable from human output. These capabilities are underpinned by advanced \ac{DL} architectures, leveraging powerful neural frameworks such as transformers and  discriminator-based models to learn and sample from complex data distributions. Through stochastic generation mechanisms, these architectures support high output diversity and contextual adaptability, even from identical prompts. \ac{GenAI} systems are trained using unsupervised, semi-supervised, and supervised learning paradigms, allowing them to generalise across modalities and domains~\cite{kasneci2023chatgpt}. 
This design evolution marks a paradigm shift in human–AI collaboration enables the automation of creative and cognitive processes once considered exclusive to human intelligence. As a foundational pillar of next-generation \ac{AI}, \ac{GenAI} is poised to transform sectors spanning education, scientific research, healthcare, and the creative industries.
Central to GenAI’s rise are several core model families--Autoregressive Models \cite{larochelle2011neural}, \acp{VAE} \cite{kingma2013auto}, \acp{GAN} \cite{goodfellow2014generative}, Flow-based Models and Diffusion Models. Each contributes unique mechanisms for learning, representation, and sample synthesis: from probabilistic factorisation and latent variable modeling to adversarial loss dynamics and iterative noise-based refinement. 
Their ongoing advancement has directly influenced the design of \acp{LLM}, which now constitute a dominant paradigm within the broader \ac{GenAI} ecosystem. These models embody the convergence of scale, representation learning, and semantic modeling, positioning them as the foundation of modern AI capabilities. 


A major breakthrough in generative modelling came with the introduction of the Transformer architecture, which replaces recurrence with self-attention mechanisms. This innovation enabled parallel training and improved the modelling of long-range dependencies, forming the backbone of modern \acp{LLM}. 
\acp{LLM}, such as GPT-2 and GPT-3 are transformer-based models trained on large, diverse corpora to perform language understanding and generation. 
Their scaling from millions to hundreds of billions of parameters has yielded capabilities like few-shot and zero-shot generalisation \cite{ouyang2022training}. 


\subsection{Challenges in Generative AI and LLMs}

Despite their transformative capabilities, \ac{GenAI} and \acp{LLM} face significant challenges spanning technical, practical, and ethical domain.
A fundamental issue is their dependence on vast, diverse datasets for training. Achieving state-of-art performance typically necessitates access to large-scale corpora, often containing sensitive or personally identifiable information. This raises substantial concerns regarding data privacy, consent, data ownership, and risk of inadvertent private information leakage especially as models can sometimes reproduce fragments from their training data. To address these challenges, emerging techniques such as differential privacy, \ac{FL}, and model distillation are being actively explored. These approaches aim to retain model utility while safeguarding user data, promoting ethical AI development aligned with national and regulatory privacy standards.

Another pressing challenge lies in the enormous computational cost associated with training and increasingly deploying advanced generative models.  For instance, the training of GPT-3 reportedly incurred compute costs exceeding $4.6$ million, equivalent to roughly 355 \ac{GPU}-years if done naively on a single \ac{GPU} \cite{gpt3ref}. 
Although such training is distributed across high-performance cluster of \acp{GPU} or \acp{TPU} to reduce time-to-train, the financial and environmental burden remain substantial \cite{Ahmadi_sustainability25}. While optimisation strategies such as model pruning, knowledge distillation, efficient architectural designs, and algorithmic enhancements have shown promise in improving compute efficiency, frontier models continues to push the boundaries of current hardware capabilities. 
These computational demands directly translate into sustainability concern. The energy required to train and operate large-scale AI models significantly contributes to the carbon footprints, exacerbated by data centre power usage, cooling requirements, and hardware turnover. Addressing this environment footprint is imperative. Advances in low-power AI hardware, energy-aware scheduling, and green AI practices not only reduce operational costs but also align with global climate objectives and sustainable digital infrastructure goals. 

Beyond privacy and sustainability, \ac{GenAI} and \acp{LLM} raises profound concerns around interpretability, controllability, and responsible deployment. Due to their inherent complexity, comprising billions of parameters, \acp{LLM} often function as closed-boxes with limited transparency into how outputs are generated. The lack of explainability  complicates tasks as debugging, output justification, trust calibration, and regulatory compliance. Furthermore, societal risk ranging from misinformation propagation to automation bias and algorithmic manipulation, underscores the urgency of establishing robust governance framework and comprehensive risk mitigation strategies. 

Ultimately, mitigating these challenges requires an interdisciplinary approach, effort encompassing \ac{ML} research, systems engineering, privacy legislation, and sustainability policy. As \ac{GenAI} and \acp{LLM} become integral to digital infrastructure, including national telecommunications networks, the need for Sovereign \ac{AI} becomes increasingly critical. Sovereign control over data, compute infrastructure, and model governance is not only essential for privacy protection and regulatory alignment but also for ensuring trustworthy, secure, and resilient AI ecosystems that serve national strategic interests.

\section{Sovereign AI} 

In the rapidly evolving digital ecosystem, especially in telecoms businesses, organisations and other entities are increasingly developing own AI models or training \acp{LLM} with available data. This trend raises critical questions: if AI to remain open-source, who controls the underlying data? And are there instances where this data may conflict with a country's and/or organisation's culture, economic objective, or  security? This concern highlights the importance of a sovereign AI infrastructure, often refereed as sovereign AI. What once considered as a strategic asset has now become a foundational necessity. As \ac{AI} systems become deeply integrated into core society functions, including national security and critical infrastructure, economic productivity, and public services, the critical issues is not whether \ac{AI} should be adopted, but rather who governed, control, and benefits from its deployment. 

At its core, \ac{AI} sovereignty refers to the ability of governments, national industries, and institutions to independently develop, govern, and deploy AI systems in alignment with domestic laws, ethical values, and strategic priorities. This autonomy minimises dependency on foreign technologies and mitigates associated risks such as foreign surveillance, monopolistic influence, and loss of regulatory oversight. In the context of \ac{6G} and other critical infrastructure, sovereignty is imperative for preserving national security, economic resilience, and technological competitiveness.


\subsection{Why AI Sovereignty Matters}
\subsubsection{National Security and Data Protection}
The increasing reliance on AI-powered cloud services, third-party platforms, and external data centres introduces significant vulnerabilities in digital governance. \ac{AI} systems now influence core decisions in cybersecurity, defence operations, and public safety. When these systems operate on unverifiable-built platforms, the risks of surveillance, manipulation, and cyber-attacks escalate significantly. Sovereign \ac{AI} frameworks allow for the deployment of secure, locally governed \ac{AI} pipelines that safeguard sensitive citizen, corporate, and governmental data from foreign access and manipulation. By embedding \ac{AI} into nationally controlled infrastructures, states can ensure the integrity, confidentiality, and operational independence of their digital environments. 

\subsubsection{Technological Independence \& Economic Resilience}
Beyond national defence, \ac{AI} sovereignty serves as a catalyst for economic self-reliance and innovation leadership. Countries that nurture home-grown \ac{AI} ecosystems are better positioned to shape their technological futures in line with national priorities. Such ecosystems support the development of indigenous technologies, protect intellectual property, and generate high-value employment opportunities. 

Moreover, nations/organisations that maintain control over their \ac{AI} infrastructure and training datasets gain the capacity to establish global standards, rather than being subjected to external norms or regulatory constraints. Sovereign \ac{AI} thus becomes a strategic enabler for sustainable economic growth, reduced dependency on external monopolies, and enhanced competitiveness in the digital economy. 

\subsubsection{Ethical \& Regulatory Autonomy}
Different nations possess unique ethical frameworks, legal traditions, and social values. The use of foreign-developed AI systems risks misalignment with domestic norms, particularly when algorithmic decisions are governed by opaque logic or embedded with external biases. Sovereign \ac{AI} enables the enforcement of local governance standards, supports transparency and accountability in AI decision-making, and ensures compliance with data protection laws such as the GDPR or equivalent national regulations. Furthermore, it facilitates the creation of culturally sensitive and contextually appropriate AI systems that reflect local societal norms, languages, and policy goals.

\subsection{Resilient Digital Ecosystems \& Infrastructure}
The resilience of a nation’s digital infrastructure is critically linked to its degree of autonomy in managing \ac{AI}-enabled networks, services, and applications. A sovereign \ac{AI} ecosystem ensures that core digital services are not susceptible to non-sovereign control, technological closed boxes, or extraterritorial legal mandates. 
In the context of next-generation communication systems like \ac{6G}, sovereign \ac{AI} enables intelligent automation of radio access and core network functions such as spectrum optimisation, resource scheduling, and network security without reliance on external control planes. Furthermore, by designing interoperable frameworks, sovereign \ac{AI} systems can maintain global compatibility while preserving autonomy, enabling secure and standards-compliant international collaboration.

\subsection{Challenges and drawbacks}
Developing and maintaining sovereign \ac{AI} have challenges and drawbacks like high development and deployment costs. This increases the consumer price and affects the competitiveness. This can be similar to the case that in some geographical areas \acp{MNO} deploy their own infrastructure and in others share the infrastructure to mange the \ac{CAPEX} and \ac{OPEX}. This will be significantly more complicated for \ac{AI}. In the case of \acp{MNO}, to overcome this challenge large multi-national operators can have a single organisational sovereign \ac{AI} operating in all areas that they have presence. The smaller, also known as tier two, \acp{MNO}, can partner with \ac{AI} and cloud providers to deploy their sovereign \ac{AI}. This middle ground approach, however, may not give them full sovereignty of their deployed \ac{AI}. An important step is forming an international consortium where countries agree on ethical data-use guidelines for AI training, ensuring interoperability and cross-border trust.



\section{AI in telecommunications networks}




\subsection{Modern AI}
In \ac{6G}, \ac{AI}-native design will be foundational. The network will not merely transport data but interpret, reason, and autonomously act upon the information flows it carries. Several works have already discussed different approaches of integration of \ac{AI} in \ac{6G} using different techniques \cite{sun_xAI25}. In this work, due to the lack of space we only focus on more modern and complex forms of \ac{AI} in \ac{6G}.

\subsubsection{Generative AI}\ac{GenAI} models are positioned to become instrumental across the \ac{6G} ecosystem. They will enable dynamic service composition, real-time network slicing configuration, predictive traffic management through synthetic environment modelling and data augmentation. By generating  plausible scenarios for network behaviours under varying load, interference, or mobility patterns, \ac{GenAI} framework will significantly enhance network resilience and adaptive capacity. 

\subsubsection{Large Telecom Models}Alongside \ac{GenAI}, \acp{LLM}, initially deployed for natural language tasks, and are increasingly adapted to network control and management tasks. Within \ac{6G} systems, \acp{LLM} can mediate between human operators and complex network functions, translating high-level service intents into machine-executable policies and configurations through prompt-driven orchestration. \acp{LLM} facilitate intent-based networking, where user or operator intents are autonomously fulfilled by the network in real time. 

Further specialisation is evident in the emergence of \acp{LTM}--telecommunication-specific \acp{LLM} fine-tuned on domain-specific corpora, such as mobility pattern, spectrum allocation, \ac{RAN} configuration and \ac{SLA} parameters. \acp{LTM} enable precise, context-aware optimisation including dynamic handover decision-making, energy management, and fault detection in disaggregated and virtualised network slices. Moreover, by embedding telecom-native semantics, \acp{LTM} reduce risk and improve trustworthiness compared to general-purpose \acp{LLM} \cite{LTM_paper}. 

\subsubsection{AI at the Edge} The nature of \ac{6G} infrastructure accentuates the role of Edge \ac{AI}. Real-time inference for mission-critical applications such as autonomous driving, telesurgery, or immersive \ac{XR} requires \ac{AI} models to be executed at the network edge with stringent latency, reliability, and privacy constraints. Edge-deployed \ac{AI} agents will dynamically manage local radio resources, perform context-aware beamforming, predict user mobility, and pre-emptively optimise service provisioning without reliance on centralised control.

\subsubsection{Agentic AI} Beyond traditional inference and optimisation tasks, \ac{6G} networks will increasingly require Agentic \ac{AI} systems intelligent agents capable not only of executing specific models but also of autonomously formulating multi-step strategies, adapting to unforeseen environmental conditions, and achieving operator-defined high-level goals with minimal human intervention. Within telecom infrastructure, Agentic functions will dynamically compose service chains, optimise end-to-end \ac{QoS} across heterogeneous radio and core networks, and proactively adjust network configurations in response to emerging traffic patterns, interference dynamics, or security threats. Unlike traditional model-driven approaches, Agentic \ac{AI} in \ac{6G} must integrate planning, reasoning, and semantic understanding into the control loops of the intelligence controllers, edge nodes, and even user-facing services. This transition fundamentally elevates the need for auditable autonomy, verifiable decision chains, and sovereign governance over \ac{AI} lifecycles to maintain operational integrity and strategic independence.

\subsection{Challenges of Modern AI in 6G}
The increasing autonomy and decentralisation of \ac{AI} models across the network fabric introduce significant challenges in terms of data protection, trustworthiness, and control. AI agents embedded at the \ac{RAN}, core, and edge levels will access sensitive operational data, user mobility traces, and encrypted payloads, creating new surfaces for privacy violations and adversarial exploitation. Furthermore, training, updating, and governing AI models across a heterogeneous, multi-vendor, and often geopolitically sensitive environment requires strict compliance with data localisation laws, secure model lifecycle management, and verifiable accountability mechanisms.  

The stakes are further elevated by the fact that in \ac{6G}, network performance, security, and even business-critical service levels will be determined by autonomous \ac{AI} decisions. This case becomes even more serious for the case of private and mission critical network operators. They require trustable \ac{AI} to operate on their network. However, due to their smaller business size they are not capable of maintaining their own \ac{LTM}.

Without sovereign governance over the AI models controlling their training datasets, architectures, update cycles, and inference behaviours operators and nations risk ceding control over critical digital infrastructure to opaque, external entities.


\section{Sovereign AI for 6G and Open RAN Networks}

As \ac{AI} becomes foundational to the real-time operation, orchestration, and strategic decision-making within \ac{6G} networks, the imperative for sovereign control over \ac{AI} models, data pipelines, and infrastructure becomes critical. 
Sovereign \ac{AI} for \ac{6G} networks refers as a nation's (in our case operator's) capability to develop, deploy, manage, and regulate \ac{AI} technologies across the ecosystem. This concept extends significantly beyond traditional data sovereignty, which primarily focuses on legal and regulatory control over data based on location. Sovereign \ac{AI} encompasses the full stack and lifecycle: secure and institutionally controlled compute infrastructure (including edge and cloud), sovereign data governance frameworks should be compliant with local regulations, verifiable and trusted algorithms (including foundation models and specialised agents), the development of a domestic \ac{AI} talent pool, and robust governance frameworks aligned with national interests, cultural values, security requirements, and legal structures (e.g., GDPR, national data protection acts).

\acp{MNO} with their existing in-region infrastructure, spectrum assets, regulatory expertise, and established relationships, are uniquely positioned to become key enablers and beneficiaries of Sovereign AI, potentially moving up the value chain from connectivity providers to trusted service enablers for government and enterprise AI solutions. 


\subsection{Ensuring AI Safety, Explainability and Ethical Integrity}
\ac{AI}-native 6G networks, where complex, high-impact decisions such as dynamic network slicing resource allocation, predictive beam management in cell-free MIMO, autonomous cyber-threat response, or intent-based network orchestration are increasingly delegated to \ac{AI} agents, necessitates a paradigm shift towards embedding Trustworthy \ac{AI} principles by design. 
This delegation occurs within evolving network functions, including those within the \acp{RIC} of \ac{O-RAN} architectures and core network functions like the \ac{NWDAF}, where \ac{AI} governs real-time decisions across latency-sensitive and mission-critical domains.

\ac{AI} models in \ac{6G} are susceptible to a range of sophisticated threats. Data poisoning can occur during collaborative training, particularly in \ac{FL} environments like \ac{RIC}-based training of xApps and rApps, where malicious updates may degrade model performance or introduce backdoors. Evasion attacks exploit inference-time weaknesses by crafting subtle perturbations in network metrics, deceiving \ac{AI} models in functions such as intrusion detection or beam allocation. Model stealing and inference attacks further undermine data privacy, exposing sensitive information or intellectual property. Moreover, \ac{GenAI} threats enabled by large-scale \acp{LLM} or diffusion models can generate synthetic yet harmful inputs, fabricate network telemetry, or create misleading policies and configurations. 

To safeguard \ac{6G} infrastructure against these risks, Sovereign \ac{AI} frameworks must institutionalise advanced defence mechanisms. Adversarial training, where models are exposed to malicious perturbations during learning, enhances robustness but requires continual adaptation. Input validation and sanitization are essential at \ac{RAN} and edge interfaces to filter anomalous \ac{KPI} formats or corrupted signal features. Certified defences such as model randomisation introduce stochasticity that complicates adversarial probing, while defensive distillation and bounded model checking provide verification guarantees within bounded complexity. In \ac{FL} contexts, robust aggregation mechanisms are vital. Methods such as the geometric median or Krum ensure that Byzantine or poisoned updates do not corrupt the global model. These mechanisms are particularly relevant to \ac{AI} lifecycle orchestration in distributed O-RAN environments. The architectural flow and integration points of Sovereign AI across Non-RT and Near-RT RIC components are illustrated in 


Ultimately, \ac{AI} safety and ethical compliance in sovereign 6G environments is not an adjunct concern, it is the backbone of trusted autonomy. Ensuring that models deployed across heterogeneous, multi-vendor, and geo-politically diverse infrastructure can resist attack, offer interpretability, and conform to value-based regulations is the definitive prerequisite for a resilient, inclusive, and future-proof \ac{AI}-native network.

\subsection{AI-Powered Troubleshooting and Operational Intelligence}
Sovereign AI frameworks enable the deployment of telecom-specific AI troubleshooting agents trained on local network telemetry, KPIs, and incident datasets. These agents, operating within \ac{Near-RT} and \ac{Non-RT} \ac{RIC} domains, must support fault prediction, diagnosis, and recovery under strict privacy and compliance constraints.
Unlike generic \ac{LLM} copilots, sovereign telecom copilots must incorporate policy-constrained inferencing, zero-trust model updates, and auditable decision paths to guarantee security and maintain operator autonomy. \ac{FL} approaches at the \ac{RIC} level can ensure collaborative model improvements without leaking sensitive network data across operators or borders.

\subsection{AI for R\&D, Local Innovation, and Strategic Autonomy}
Strategic autonomy in the context of \ac{6G} requires more than infrastructural control; it demands the ability to foster indigenous \ac{AI} innovation tailored to national priorities. At the core of this effort is a secure, federated, and sovereign AI R\&D environment capable of supporting the design, training, and deployment of AI models within national jurisdiction.

A foundational requirement is nationally hosted AI infrastructure sovereign cloud platforms and \ac{AI} factories equipped with high-performance compute nodes (e.g., GPUs and TPUs). These systems enable training of large, domain-specific models while adhering to localisation mandates. Complementary to infrastructure is the curation of large-scale, anonymised datasets encompassing wireless channel data, mobility patterns, traffic statistics, and spectrum usage sourced from national telecom networks.
To address data access limitations while upholding privacy laws, sovereign AI ecosystems increasingly rely on \acp{PET} enabling collaborative analytics without data exposure. Furthermore, \ac{SDG} becomes indispensable. Advanced generative frameworks including GANs, VAEs, and especially diffusion models support the creation of high-fidelity training datasets, aiding reinforcement learning, digital twin simulation, and fault injection testing.

National \ac{AI} testbeds and federated sandboxes play a pivotal role in enabling controlled experimentation with xApps, rApps, and sovereign \acp{LLM} under real-time constraints. Regulatory sandboxes governed by national authorities further permit pre-commercial validation of \ac{AI} functions within legal frameworks. Sovereign Digital Twins, representing virtual replicas of \ac{6G} infrastructure, facilitate simulation of network conditions and ``what-if” analyses, dramatically reducing real-world risk.
Collaboration between telecom operators, universities, and research labs must focus on mission-critical domains including cell-free MIMO optimisation, semantic communication for regional languages, energy-aware resource allocation, and \ac{AI}-native PHY layer designs. 
Open-source engagement is equally vital. Sovereign ecosystems should support and leverage trusted open-source telecom and \ac{AI} toolkits to avoid vendor lock-in, accelerate innovation, and foster community contributions under sovereign control.

Ultimately, sovereign \ac{AI} R\&D must resolve the inherent tension between the scale of data needed to build high-performing models and the constraints of national data governance. The maturity of sovereign experimentation environments, synthetic generation pipelines, and federated learning platforms will determine a nation’s ability to achieve technological and policy independence in the \ac{6G} era.



\subsection{Training the Next Generation of Telecom-AI Engineers}
The long-term viability of sovereign \ac{AI} in \ac{6G} networks hinges not only on infrastructure and governance, but on cultivating a skilled, interdisciplinary workforce capable of designing, deploying, and maintaining \ac{AI}-native telecommunications systems. As network intelligence becomes increasingly decentralised and automated, future telecom engineers must possess foundational expertise not only in radio access and core network architecture, but also in machine learning, distributed AI systems, privacy-preserving technologies, and secure system design.

To support this paradigm shift, national and operator's education and R\&D frameworks must integrate AI/ML curricula into telecommunications and computer engineering programs, ensuring that graduates are equipped to contribute meaningfully to sovereign \ac{AI} initiatives. Beyond theory, immersive hands-on environments are essential. Sovereign \ac{AI} platforms should enable real-time experimentation with xApps/rApps, intent-driven orchestration logic, and \ac{FL} workflows using simulated or synthetic network environments.
These platforms must also support experimentation with data governance models, explainable \ac{AI} frameworks, and secure inference protocols under realistic \ac{6G} operational constraints. Integration of digital twins and domain-specific \ac{GenAI} copilots can further enhance learning by simulating complex network behaviours and AI decision chains in a controlled environment.

Public-private partnerships play a critical role in creating a sustainable talent pipeline. Government-backed AI centres of excellence, industrial PhD programs, and collaborative R\&D initiatives can bridge the gap between academia and sovereign telecom industry needs. These efforts must align with national strategies for digital sovereignty, ensuring that the workforce is not only technically competent but also grounded in the ethical, legal, and geopolitical considerations that underpin sovereign \ac{AI} deployments.

Without a robust pool of \ac{AI}-literate telecom engineers, efforts to operationalise sovereign \ac{AI} in \ac{6G} will face significant bottlenecks. Training the next generation of engineers must therefore be recognised as a strategic imperative equal in importance to investments in data infrastructure or algorithmic.

\begin{figure}
    \centering
    \includegraphics[width=0.8\columnwidth]{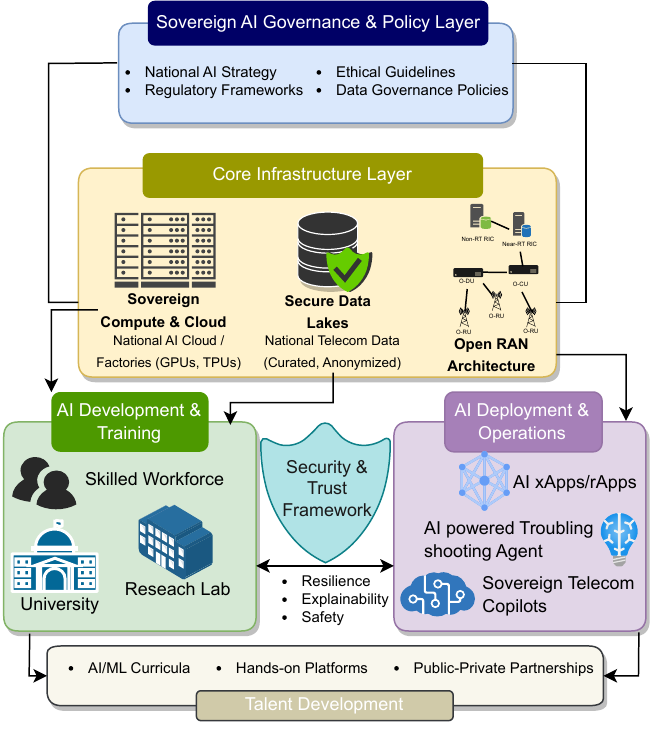}
    \caption{ Integrated Sovereign AI Ecosystem for 6G and O-RAN}
    \label{fig:blueprint}
\end{figure}

Figure \ref{fig:blueprint} depicts a blueprint for Sovereign AI in 6G networks involves a multi-faceted approach (as described above), integrating architectural, operational, and governance frameworks to ensure national or operator-level control over AI development, deployment, and lifecycle management.  This is particularly critical as telecom networks transition to AI-native 6G architectures, moving beyond 5G's cloud-centric model. 

\section{The Global Landscape: Sovereign AI Initiatives}

The  concept of Sovereign \ac{AI} is rapidly translating into concrete national strategies policies, and industry initiative across the globe. While approaches vary based on national priorities, resources and geopolitical contexts, a clear trend towards asserting greater control over \ac{AI} capabilities, particularly in telecommunications section is evident.

\newcolumntype{Y}{>{\centering\arraybackslash}X}
\begin{table*}[t]
\centering
\small
\renewcommand{\arraystretch}{1.1}
\footnotesize
\begin{tabularx}{\textwidth}{|>{\raggedright\arraybackslash}X|Y|Y|Y|Y|Y|Y|}
\hline
\rowcolor{gray!15}
\textbf{Feature} & \textbf{EU} & \textbf{India} & \textbf{Canada} & \textbf{Singapore} & \textbf{US}& \textbf{UK}\\
\hline
\textbf{Key Policy Driver(s)} & Digital Sovereignty, Privacy (GDPR), Economy, Values & Self-Reliance, Economic Growth, Local Languages, Security & Innovation, Talent, Economic Growth & Regional Relevance, Values, Economy & National Security, Economic Competition & Economic Growth, National Security, \\
\hline
\textbf{Infrastructure Focus} & Gaia-X, Sovereign Cloud, Edge & Compute Infra, Public-Private Partnerships & AI Factories, SME Support & Local AI Ecosystem, Private Leverage & Domestic Compute, Chip Control & AIRR, AI Growth Zones, National Supercomputing\\
\hline
\textbf{Data Strategy} & GDPR, Interoperability & National Datasets, Bhashini & Shared Access, Talent Focus & SEA-LION (LM) & Data/Model Control & Secure Data Infra \& Localisation\\
\hline
\textbf{Telco Role} & User/Partner in Gaia-X & Infra Partner, Service Enabler & Infra Provider & Ecosystem Partner & Critical Infra User & Critical Infra, Digital Safety\\
\hline
\textbf{Key Programs} & Gaia-X, AI Act, SNS JU (6G) & AI Strategy, Bhashini, Tata-NVIDIA & AI Strategy, TELUS Factories & SEA-LION Initiative & Export Controls, Strategic AI Support &  AI Safety Institute, AIRR\\
\hline
\textbf{Challenges} & Foreign Tech Dependence, Fragmentation & Infra Scaling, Data Diversity & Talent Gaps, Scaling Capacity & Local vs. Global Trade-off & Global Control vs. Sovereignty & Talent, Investment, Energy, Openness vs. Control\\
\hline
\end{tabularx}
\caption{Sovereign AI Approaches Across Leading Nations}
\label{tab:sovereign-ai-comparison}
\end{table*}

\subsection{National Strategies and Implementation Models}
As the demand for AI-native infrastructure grows, several nations are translating the concept of Sovereign \ac{AI} into concrete strategies, supported by the real-world implementation. While national approaches vary based on geopolitical context, priorities, and technological maturity, common themes include local data control, infrastructure autonomy, regulatory alignment, and economic self-reliance. 

\subsubsection{Europe (EU)} The EU has placed digital sovereignty high on its agenda, driven by concerns over dependence on US and Chinese tech giants, data privacy (GDPR), and economic competitiveness. The strategy emphasises regulation (e.g., AI Act, Digital Markets Act), fostering a European data infrastructure (Gaia-X \cite{gaiax2021}), and promoting AI aligned with European values. Gaia-X aims to create a federated, secure, and interoperable data ecosystem based on principles of transparency and user control, enabling sovereign cloud services. The Smart Networks and Services Joint Undertaking (SNS JU) is funding \ac{6G} research explicitly linked to strengthening Europe's technological sovereignty and leadership in standards. As key EU members, France and Germany are strong proponents of digital sovereignty, actively supporting initiatives like Gaia-X and promoting national cloud providers (e.g., OVHcloud, T-Systems) that comply with European standards and aim to reduce reliance on non-EU technologies.
\subsubsection{United Kingdom} The UK has similarly committed to a robust sovereign AI agenda, albeit with a distinct approach that balances national control with an open, innovation-friendly environment. Driven by both economic ambition and the need to secure critical technological capabilities, the UK's strategy focuses on building substantial domestic compute power (e.g., the AI Research Resource (AIRR \cite{ukSovereignAI2024}) and supercomputers like Isambard-AI), fostering a vibrant AI ecosystem through initiatives like AI Growth Zones to attract investment in data centres, and establishing the AI Energy Council to ensure the energy infrastructure can support this growth. The creation of a dedicated Sovereign AI Unit underscores the government's commitment to orchestrating these efforts, ensuring the UK can develop, deploy, and control its AI in line with its national interests, while its ``pro-innovation" regulatory stance aims to promote safety without stifling the rapid pace of AI advancement.
\subsubsection{Asia} India is pursuing a comprehensive national AI strategy focused on inclusive growth and technological self-reliance. Key initiatives include the Bhashini platform for developing AI translation tools in diverse Indian languages, establishing AI standards for ethical practices, and fostering partnerships to build significant domestic AI infrastructure, exemplified by collaborations between Indian conglomerates (Reliance, Tata) and NVIDIA. There are reports of ambitious plans for large-scale AI compute infrastructure build-outs. Moreover, Singapore focuses on developing AI capabilities tailored to its regional context, including the development of the SEA-LION \acp{LLM} \cite{ng2025sealion} designed to better represent Southeast Asian languages and cultures. The nation is actively designing a sovereign AI ecosystem leveraging private sector expertise. 

\subsubsection{Canada} Canada's Pan-Canadian AI Strategy emphasises research, talent development, and support for \acp{SME}. There is also movement in the private sector, with major telecom operator TELUS announcing plans to launch Sovereign AI factories to bolster digital sovereignty and innovation.

   
Table~\ref{tab:sovereign-ai-comparison} provides a comparative overview of selected national approaches. A notable dynamic emerging from these initiatives is the heavy reliance on public-private partnerships and collaborations with major international technology vendors, particularly for providing the advanced compute hardware (\acp{GPU}) and \ac{AI} software platforms needed. While these partnerships accelerate the deployment of sovereign infrastructure, they introduce a potential tension with the core objective of reducing dependency on foreign entities. Nations pursuing Sovereign AI may find themselves navigating a complex landscape where they achieve localised data control and governance but remain strategically reliant on the ecosystems and technological roadmaps of a few dominant global players, potentially trading one form of dependency for another. Understanding and managing these nuanced dependencies will be crucial for the long-term success of national Sovereign AI strategies.

\subsection{Industry Movements \& Collaborations}

\subsubsection{Industry Alliances} Groups like AI-RAN Alliances are emerging, and bring together industry players to focus on integrating AI into \ac{RAN} for 5G and 6G. While focused on technology development, these alliances inevitably intersect with sovereignty concerns and may influence future standards and deployment models.

\subsubsection{Vendor Partnerships} Building sovereign AI capabilities often involves collaboration. Major technology vendors like NVIDIA, Dell, Vertiv, Wipro and cloud provider like AWS are actively partnering with national government and local telecom operators to provide the necessary hardware (GPUs, servers), software platform, and expertise to build and operate sovereign AI infrastructure. Like, Swisscom partnering with NVIDIA in Switzerland, Indosat Ooredoo Hutchison collaborating with NVIDIA and others in Indonesia and Wipro offering sovereign AI services globally using NVIDIA technology. 

\subsubsection{Operator AI Deployments} Research firms like Juniper track AI deployments by major global operators (e.g., AT\&T, China Mobile, Deutsche Telekom, Vodafone). While not always explicitly labeled ``sovereign," these deployments increasingly incorporate considerations of data localisation, security, and compliance relevant to sovereignty.

\subsubsection{Specific Projects} The Wipro/SIAM.AI collaboration for the Tourism Authority of Thailand and the iGenius/Vertiv/NVIDIA sovereign AI data center project in Italy provide concrete examples of sovereign AI implementations targeting specific sectors or regional needs.

The global landscape of Sovereign \ac{AI} reveals a shared urgency among nations to assert greater control over data, infrastructure, and \ac{AI} capabilities particularly within the telecommunications domain. While national strategies differ in scope and maturity, the convergence of regulatory frameworks, public-private partnerships, and localised \ac{AI} initiatives signals a collective shift toward \ac{AI} autonomy. As case studies demonstrate, operationalising sovereignty is both technically feasible and strategically necessary. Going forward, the ability to balance local control with international collaboration will be pivotal in shaping resilient and inclusive AI ecosystems for the 6G era.

\section{Conclusion and Future directions}

The transition to \ac{6G} networks demands a rethinking of infrastructure design, model lifecycle management, and regulatory enforcement across the telecom stack. As foundational AI models such as \ac{GenAI}, \acp{LLM} and agentic models becomes central to network orchestration, control plane automation, and service composition, the need for operational sovereignty becomes paramount. Sovereign AI addresses this need by enabling end-to-end ownership and control over AI pipeline, including data ingestion, model training, inference and policy compliance and auditability within an operator's domain.  

Embedding Sovereign AI within components such as the Near-RT and Non-RT RICs in O-RAN architectures, and extending it to edge-cloud continuum environments, ensures that AI agents operate under trusted, explainable, and policy-aligned conditions. It also mitigates dependency on foreign-controlled AI stacks, thereby securing infrastructure against adversarial threats, compliance violations, and unauthorised model behaviour. 

Operators and national stakeholders must invest in modular, federated, and privacy-preserving AI frameworks backed by sovereign compute infrastructure, national data assets, and vertically integrated governance. This technical sovereignty will be foundational not only for secure and resilient telecom operations, but also for enabling future government and enterprise services in a highly distributed, AI-driven 6G ecosystem.

\input{Acron}
\bibliographystyle{IEEEtran}

\bibliography{References}

\end{document}

%% file: Acron.tex
\begin{acronym} 

\acro{3GPP}{Third Generation Partnership Project}
\acro{3D-InteCom}{Three-Dimensional Integrated Communications}
\acro{5G}{The fifth generation of mobile networks}
\acro{6G}{Sixth generation of mobile networks}
\acro{ACO}{Ant Colony Optimization}
\acro{AI}{Artificial Intelligence}
\acro{AR}{Augmented Reality}
\acro{ANN}{Artificial Neural Network}
\acro{AP}{access point}
\acro{API}{Application Programming Interface}
\acro{BB}{Base Band}
\acro{BBU}{Base Band Unit}
\acro{BER}{Bit Error Rate}
\acro{BS}{Base Station}
\acro{BW}{Bandwidth}
\acro{BIRCH}{Balanced Iterative Reducing and Clustering using Hierarchies}
\acro{BE}{Best-Effort}
\acro{BigCom}{Big Communications}
\acro{CAV}{Connected and Autonomous Vehicle}
\acro{CC}{Chain Composition}
\acro{CF}{Cell-Free}
\acro{C-RAN}{Cloud Radio Access Networks}
\acro{CAPEX}{Capital Expenditure}
\acro{CoMP}{Coordinated Multipoint}
\acro{COTS}{Commercial-Off-The-Shelf}
\acro{CR}{Cognitive Radio}
\acro{CU}{Central Unit}
\acro{COC}{Computation Oriented Communications}
\acro{CAeC}{Contextually Agile eMBB Communications}
\acro{CBQ}{Class-Based Queueing}
\acro{CN}{Core Network}
\acro{D2D}{Device-to-Device}
\acro{DA}{Digital Avatar}
\acro{DAC}{Digital-to-Analog Converter}
\acro{DAS}{Distributed Antenna Systems}
\acro{DBA}{Dynamic Bandwidth Allocation}
\acro{DNN}{Deep Neural Network}
\acro{DC}{Duty Cycle}
\acro{DyPr}{`Dynamic Prioritization'}
\acro{DL}{Deep Learning}
\acro{DSA}{Dynamic Spectrum Access}
\acro{DT}{Digital Twin}
\acro{DRL}{Deep Reinforcement Learning}
\acro{DRN}{Deep Reinforcement Network}
\acro{DQL}{Deep Q Learning}
\acro{DDQL}{Double Deep Q Learning}
\acro{DTs}{Decision Tress}
\acro{DDPG}{Deep Deterministic Policy Gradient}
\acro{DPI}{Deep Packet Inspection}
\acro{$E^2D^2PG$}{Enhanced Exploration Deep Deterministic Policy}
\acro{DU}{Distributed Unit}
\acro{EE}{Energy Efficiency}
\acro{e-waste}{Electronic waste}
\acro{ER}{Erdős-Rényi}
\acro{EUB}{Expected Upper Bound}
\acro{EDuRLLC}{Event Defined uRLLC}
\acro{EVNFP}{Elastic Virtual Network Function Placement}
\acro{eMBB}{enhanced Mobile Broadband}
\acro{eMBB-Plus}{Enhanced Mobile Broadband Plus}
\acro{eCPRI}{evolved CPRI}
\acro{FBMC}{Filterbank Multicarrier}
\acro{FEC}{Forward Error Correction}
\acro{FG} {Forwarding Graph}
\acro{FGE}{FG Embedding}
\acro{FIFO}{First-in-First-out}
\acro{FCFS}{First Come, First Served}
\acro{FFR}{Fractional Frequency Reuse}
\acro{FL}{Federated Learning}
\acro{FSO}{Free Space Optics}
\acro{fBm}{Fractional Brownian motion}
\acro{GA}{Genetic Algorithms}
\acro{GenAI}{Generative Artificial Intelligence}
\acro{GAN}{Generative Adversarial Network}
\acro{gNB}{Next Generation Node B}
\acro{GI}{Granularity Index}
\acro{GM}{Gaussian Mixture}
\acro{GPP}{General-Purpose Processor}
\acro{GPU}{Graphics Processing Unit}
\acro{HAP}{High Altitude Platform}
\acro{HD}{High-Demand}
\acro{HL}{Higher Layer}
\acro{HARQ}{Hybrid-Automatic Repeat Request}
\acro{ICT}{Information and Communication Technology}
\acro{IoE}{Internet of Everything}
\acro{IoT}{Internet of Things}
\acro{ILP}{Integer Linear Program}
\acro{IRS}{Intelligent Reflective Surfaces}
\acro{ISAC}{Integrated Sensing and Communication}
\acro{ITU}{International Telecommunication Union}
\acro{KPI}{Key Performance Indicator}
\acro{KPM}{Key Performance Measurement}
\acro{KNN}{K-Nearest Neighbour}
\acro{LAN}{Local Area Network}
\acro{LAP}{Low Altitude Platform}
\acro{LC}{line card}
\acro{LEO}{Low Earth Orbit}
\acro{LL}{Lower Layer}
\acro{LLM}{Large Language Model}
\acro{LR}{Logistic Regression}
\acro{LOS}{Line of Sight}
\acro{LTE}{Long Term Evolution}
\acro{LTE-A}{Long Term Evolution Advanced}
\acro{LRD}{Long-Range Dependence}
\acro{LFGL}{Least-First-Greatest-Last}
\acro{LTM}{Large Telecom Model}
\acro{MAC}{Medium Access Control}
\acro{MAP}{Medium Altitude Platform}
\acro{MANO}{Management and Orchestration}
\acro{MIMO}{Multiple Input Multiple Output}
\acro{ML}{Machine Learning}
\acro{MME}{Mobility Management Entity}
\acro{mmWave}{millimeter Wave}
\acro{MNO}{Mobile Network Operator}
\acro{MR}{Mixed Reality}
\acro{MTC}{Machine Type Communications}
\acro{MILP}{ Mixed-Integer Linear Program}
\acro{MDP}{Markov Decision Process}
\acro{mMTC}{massive Machine Type Communications}
\acro{NAI}{Network Availability Index}
\acro{NASA}{National Aeronautics and Space Administration}
\acro{NAT}{Network Address Translation}
\acro{NHD}{Not-so-High-Demand}
\acro{NHN}{Neutral Host Network}
\acro{NN}{Neural Network}
\acro{NF}{Network Function}
\acro{NFP}{Network Flying Platform}
\acro{NTN}{Non-terrestrial networks}
\acro{NFV}{Network Function Virtualization}
\acro{NWDAF}{Network Data Analytics Function}
\acro{NS}{Network Service}
\acro{Near-RT}{Near-Real-Time}
\acro{Non-RT}{Non-Real-Time}
\acro{OFDM}{Orthogonal Frequency Division Multiplexing}
\acro{OAM}{Orbital Angular Momentum}
\acro{OPEX}{Operational Expenditure}
\acro{OSA}{Opportunistic Spectrum Access}
\acro{O-RAN}{Open Radio Access Network}
\acro{ONU}{optical network unit}
\acro{OLT}{optical line terminal}
\acro{PCA}{Principal Component Analysis}
\acro{PAM}{Pulse Amplitude Modulation}
\acro{PAPR}{Peak-to-Average Power Ratio}
\acro{PGW}{Packet Gateway}
\acro{PHY}{physical layer}
\acro{PPML}{Privacy-Preserving ML}
\acro{PSO}{Particle Swarm Optimization}
\acro{PT}{Physical Twin}
\acro{PU}{Primary User}
\acro{Pr}{Premium}
\acro{PET}{Privacy-Enhancing Technology}
\acro{QAM}{Quadrature Amplitude Modulation}
\acro{QoE}{Quality of Experience}
\acro{QoS}{Quality of Service}
\acro{QPSK}{Quadrature Phase Shift Keying}
\acro{QL}{Q-Learning}
\acro{RA}{Resource Allocation}
\acro{RAN}{Radio Access Network}
\acro{RF}{Random Forest}
\acro{RN}{Remote Node}
\acro{RRH}{Remote Radio Head}
\acro{RRC}{Radio Resource Control}
\acro{RRU}{Remote Radio Unit}
\acro{RU}{Radio Unit}
\acro{RL}{Reinforcement Learning}
\acro{RR}{Ridge Regression}
\acro{RT}{Real Time}
\acro{RIS}{Reconfigurable Intelligent Surfaces}
\acro{RIC}{RAN Intelligent Controller}
\acro{RLHF}{Reinforcement Learning with Human Feedback}
\acro{SBA}{Service Based Architecture}
\acro{SCH}{Scheduling}
\acro{SLAs}{Service Level Agreements}
\acro{SU}{Secondary User}
\acro{SCBS}{Small Cell Base Station}
\acro{SDN}{Software Defined Network}
\acro{SFC}{Service Function Chaining}
\acro{SLA}{Service Level Agreement}
\acro{SNR}{Signal-to-Noise Ratio}
\acro{SON}{Self-Organising Network}
\acro{SAR}{Service Acceptance Rate}
\acro{SE}{spectral efficiency}
\acro{SVM}{Support Vector Machine}
\acro{SLFL}{Simple Lazy Facility Location}
\acro{SURLLC}{Secure Ultra-Reliable Low-Latency Communications}
\acro{SLM}{Small Language Model}
\acro{SMC}{Secure Multi-Party Computation}
\acro{SDG}{Synthetic Data Generation}
\acro{SME}{small and medium-sized enterprise}
\acro{SMO}{Service Management and Orchestration}
\acro{TDD}{Time Division Duplex}
\acro{TD-LTE}{Time Division LTE}
\acro{TDM}{Time Division Multiplexing}
\acro{TDMA}{Time Division Multiple Access}
\acro{TL}{Transfer Learning}
\acro{TPU}{Tensor Processing Unit}
\acro{TWDM-PON}{ time- and wavelength-division multiplexed passive optical network}
\acro{TWT}{Threshold Waiting Time}
\acro{THz}{sub-Terahertz} 
\acro{UE}{User Equipment}
\acro{UAV}{Unmanned Aerial Vehicle}
\acro{URLLC}{Ultra-Reliable Low Latency Communications}
\acro{USRP}{Universal Software Radio Platform}
\acro{UCDC}{Unconventional Data Communications }
\acro{UHD}{Ultra-High Definition}
\acro{VL}{Virtual Link}
\acro{VNF}{Virtual Network Function}
\acro{VNF-FG}{VNF-Forwarding Graph}
\acro{VNF-FGE}{VNF-FG Embedding}
\acro{VM}{Virtual Machine}
\acro{VR}{Virtual Reality}
\acro{VAE}{Variational Autoencoder}
\acro{WFQ}{Weighted Fair Queuing}
\acro{WDM MUX}{WDM multiplexer}
\acro{XAI}{eXplainable Artificial Intelligence}
\acro{XR}{eXtended Reality}
\acro{V2X}{vehicle-to-everything}
\end{acronym}

%% file: main.bbl
\begin{thebibliography}{10}
\providecommand{\url}[1]{#1}
\csname url@samestyle\endcsname
\providecommand{\newblock}{\relax}
\providecommand{\bibinfo}[2]{#2}
\providecommand{\BIBentrySTDinterwordspacing}{\spaceskip=0pt\relax}
\providecommand{\BIBentryALTinterwordstretchfactor}{4}
\providecommand{\BIBentryALTinterwordspacing}{\spaceskip=\fontdimen2\font plus
\BIBentryALTinterwordstretchfactor\fontdimen3\font minus \fontdimen4\font\relax}
\providecommand{\BIBforeignlanguage}[2]{{%
\expandafter\ifx\csname l@#1\endcsname\relax
\typeout{** WARNING: IEEEtran.bst: No hyphenation pattern has been}%
\typeout{** loaded for the language `#1'. Using the pattern for}%
\typeout{** the default language instead.}%
\else
\language=\csname l@#1\endcsname
\fi
#2}}
\providecommand{\BIBdecl}{\relax}
\BIBdecl

\bibitem{andrews_5G_2014}
J.~G. Andrews, S.~Buzzi, W.~Choi, S.~V. Hanly, A.~Lozano, A.~C. Soong, and J.~C. Zhang, ``{What will 5G be?}'' \emph{IEEE Journal on selected areas in communications}, vol.~32, no.~6, pp. 1065--1082, 2014.

\bibitem{Ahmadi_DT_21}
H.~Ahmadi, A.~Nag, Z.~Khan, K.~Sayrafian, and S.~Rahardja, ``{Networked Twins and Twins of Networks: An Overview on the Relationship Between Digital Twins and 6G},'' \emph{IEEE Communications Standards Magazine}, vol.~5, no.~4, pp. 154--160, 2021.

\bibitem{WhitePaper_LTM}
\BIBentryALTinterwordspacing
A.~Mokh, A.~De~Domenico, A.~Karapantelakis, C.~Huang, C.~Chaccour, F.~Abdeldayem, J.~Deng, K.~Ball, L.~Bariah, M.~Debbah, N.~Nikaein, O.~Hashash, Q.~Zhao, and S.~E. Cherrared, ``{Large-scale AI in Telecom : Charting the roadmap for innovation, scalability, and enhanced digital experiences},'' \emph{White Papaer}, 2025. [Online]. Available: \url{https://cdn.jsdelivr.net/gh/abncharts/abncharts.public.1/abnasia.org/1737392400700_compressed_www.abnasia.org.pdf}
\BIBentrySTDinterwordspacing

\bibitem{kasneci2023chatgpt}
E.~Kasneci, K.~Se{\ss}ler, S.~K{\"u}chemann, M.~Bannert, D.~Dementieva, F.~Fischer, U.~Gasser, G.~Groh, S.~G{\"u}nnemann, E.~H{\"u}llermeier \emph{et~al.}, ``{ChatGPT for good? On opportunities and challenges of large language models for education},'' \emph{Learning and individual differences}, vol. 103, p. 102274, 2023.

\bibitem{larochelle2011neural}
H.~Larochelle and I.~Murray, ``{The neural autoregressive distribution estimator},'' in \emph{Proceedings of the fourteenth international conference on artificial intelligence and statistics}.\hskip 1em plus 0.5em minus 0.4em\relax JMLR Workshop and Conference Proceedings, 2011, pp. 29--37.

\bibitem{kingma2013auto}
D.~P. Kingma, M.~Welling \emph{et~al.}, ``{Auto-encoding variational bayes},'' 2013.

\bibitem{goodfellow2014generative}
I.~J. Goodfellow, J.~Pouget-Abadie, M.~Mirza, B.~Xu, D.~Warde-Farley, S.~Ozair, A.~Courville, and Y.~Bengio, ``{Generative adversarial nets},'' \emph{Advances in neural information processing systems}, vol.~27, 2014.

\bibitem{ouyang2022training}
L.~Ouyang, J.~Wu, X.~Jiang, D.~Almeida, C.~Wainwright, P.~Mishkin, C.~Zhang, S.~Agarwal, K.~Slama, A.~Ray \emph{et~al.}, ``{Training language models to follow instructions with human feedback},'' \emph{Advances in neural information processing systems}, vol.~35, pp. 27\,730--27\,744, 2022.

\bibitem{gpt3ref}
\BIBentryALTinterwordspacing
M.~Rijmenam. (2020, Jun.) {The GPT-3 Model: What Does It Mean for Chatbots and Customer Service?} The Digital Speaker. [Online]. Available: \url{https://www.thedigitalspeaker.com/gpt-3-model-what-mean-chatbots-customer-service}
\BIBentrySTDinterwordspacing

\bibitem{Ahmadi_sustainability25}
H.~Ahmadi, M.~Rahmani, S.~B. Chetty, E.~E. Tsiropoulou, H.~Arslan, M.~Debbah, and T.~Quek, ``{Towards Sustainability in 6G and beyond: Challenges and Opportunities of Open RAN},'' \emph{IEEE Communications Standards Magazine}, pp. 1--1, 2025.

\bibitem{sun_xAI25}
H.~Sun, Y.~Liu, A.~Al-Tahmeesschi, A.~Nag, M.~Soleimanpour, B.~Canberk, H.~Arslan, and H.~Ahmadi, ``{Advancing 6G: Survey for Explainable AI on Communications and Network Slicing},'' \emph{IEEE Open Journal of the Communications Society}, vol.~6, pp. 1372--1412, 2025.

\bibitem{LTM_paper}
A.~Shahid, A.~Kliks, A.~Al-Tahmeesschi, A.~Elbakary, A.~Nikou, A.~Maatouk, A.~Mokh, A.~Kazemi, A.~Domenico, A.~Karapantelakis, B.~Cheng, B.~Yang, B.~Wang, C.~Fischione, C.~Zhang, C.~Ben~Issaid, C.~Yuen, C.~Peng, C.~Huang, and S.~Zitao, ``Large-scale ai in telecom: Charting the roadmap for innovation, scalability, and enhanced digital experiences,'' Mar. 2025, arXiv:2503.04184.

\bibitem{gaiax2021}
{Gaia-X European Association}, ``{Federated Data Infrastructure for Europe},'' \url{https://www.gaia-x.eu}, 2021.

\bibitem{ukSovereignAI2024}
{UK Government}, ``{Sovereign AI Unit: Policy Papers and Publications},'' \url{https://www.gov.uk/government/collections/sovereign-ai-unit}, 2024.

\bibitem{ng2025sealion}
\BIBentryALTinterwordspacing
R.~Ng, T.~N. Nguyen, Y.~Huang \emph{et~al.}, ``{SEA-LION: Southeast Asian Languages in One Network},'' 2025. [Online]. Available: \url{https://arxiv.org/abs/2504.05747}
\BIBentrySTDinterwordspacing

\end{thebibliography}
